New Approach to Malware Detection Using Optimized Convolutional Neural Network

Marwan Omar
Illinois Institute of Technology
Chicago, USA

**ABSTRACT** Cyber-crimes have become a multi-billion-dollar industry in the recent years. Most cybercrimes/attacks involve deploying some type of malware. Malware that viciously targets every industry, every sector, every enterprise and even individuals has shown its capabilities to take entire business organizations offline and cause significant financial damage in billions of dollars annually. Malware authors are constantly evolving in their attack strategies and sophistication and are developing malware that is difficult to detect and can lay dormant in the background for quite some time in order to evade security controls. Given the above argument, Traditional approaches to malware detection are no longer effective. As a result, deep learning models have become an emerging trend to detect and classify malware. This paper proposes a new convolutional deep learning neural network to accurately and effectively detect malware with high precision. This paper is different than most other papers in the literature in that it uses an expert data science approach by developing a convolutional neural network from scratch to establish a baseline of the performance model first, explores and implements an improvement model from the baseline model, and finally it evaluates the performance of the final model. The baseline model initially achieves 98% accurate rate but after increasing the depth of the CNN model, its accuracy reaches 99.183 which outperforms most of the CNN models in the literature. Finally, to further solidify the effectiveness of this CNN model, we use the improved model to make predictions on new malware samples within our dataset.

## I. INTRODUCTION

The use of information technology has been a blessing to modern life as it has enabled us to reach new heights in terms of how we live and work, but it has also added considerable vulnerabilities and threats to our life. An innocuous action, such as shuffling through a malicious website or opening an email attachment, can wreak havoc and disrupt the operations of modern businesses. Not routinely updating the system or unintentionally installing malicious software can completely expose a computer system to the vulnerability and risk of cyber-attacks. Cybercriminal activities have skyrocketed in recent years with hackers successfully disrupting critical operations of a sector or industry by taking an entire business organization hostage with the help of malware [1].

Ransomware is a type of malware growingly being used by cyber attackers to hold the computer system of their targets hostage until the ransom demands are fulfilled. One of the first ransomware incidents took place in 1989 when the attendees of the International AIDS conference received malware-infected floppy disks and lost their access to the files. They were then instructed to pay $189 to a specific PO Box located in Panama to get their access back to the files [22]. Today the trends of ransomware attacks have grown significantly as cyber attackers are making high-value target attacks by targeting specific organizations with significant sensitive information or financial resources. These organizations are critical to the economy of a nation like the USA. If the system of these target entities is held hostage, it affects the day-to-day operations of these critical sectors and therefore, these entities are more likely to

pay ransom to restore normal operations [22]. The 2021 ransomware attack on Colonial Pipeline, the largest refined products pipeline in the USA, is an example of such high-value target attacks. Cyber-crimes, thus, have become a multi-billion-dollar industry in recent years. Most cybercrimes/attacks involve deploying some type of malware. As antivirus technologies are becoming more robust and evolving into anti-malware software, malware creators are also coming up with more sophisticated and potent variants that are difficult to detect and can lay dormant in the background without arousing suspicion of the security controls [2].

The number of malware samples detected in the wild has been consistently growing over the past couple of years. According to research by McAfee labs, more than 1,224,628 malware threats were detected in Q4 2020, which includes a total of 7,899 unique new hashes. This necessitates bolstering the fight against malware detection and prevention especially as cyber-hackers are developing new variants of malware every day [3].

Malware classification is a crucial step for determining the name, family or type of malware, behaviors, and signatures of malware before required actions, such as removing and quarantining, can be undertaken. For malware classification, there are primarily two approaches used, including signature-based approach and behavior-based approach. Even though traditionally signature-based classification has been used more because it is precise and fast, it fails to detect malware variants produced by the application of obfuscation techniques [4], such as packing, encryption, metamorphism, and polymorphism [5]. The problem associated with signature-based classification can be resolved by the application of behavior-based classification because the behaviors of all malware variants are almost similar. However, retrieving data related to malware behaviors is time consuming because it should be collected during malware activation.

Therefore, a new approach for malware classification that has become popular in recent times is the one based on image processing [31]. It enables the classifier to detect and classify a malware by examining the image textures of the malware. Unlike the traditional signature or behavior-based malware detection classification through static or dynamic analysis, this approach does not really depend on studying the signature and behaviors of malware [31], and therefore, it overcomes some of the weaknesses associated with traditional malware classification of approaches.

Traditionally, malware detection and classification involve malware analysis, which refers to the process of monitoring the behavior and purpose of a malicious URL or file. There are three types of malware analysis used most, including static analysis, dynamic analysis, or a hybrid between the two. Static analysis focuses on retrieving features by inspecting an application's manifest and disassembled code [23]. Dynamic analysis monitors the behavior of an application during its execution while hybrid analysis monitors an application before its installation and during its execution. Of the three types of analyses, hybrid analysis is the most powerful as both static and dynamic analyses alone cannot detect the threats from the most sophisticated malware variants [23]. Xu et al (2016) in their study revealed that hybrid analysis can become more powerful if it is combined with deep learning technologies. The use of deep learning models improves the performance of malware detection and classification significantly with an accuracy of 95%-99%. [6].

Applying machine learning, particularly deep learning models, to detect malware has been around for several years and malware visualization has become a hot research topic among cyber security researchers in recent years. Different traditional machine learning approaches, including K-nearest neighbors, vector machine, random forests, decision tree, and naive bayes, have been used for the detection and classification of a known malware [24]. Malware samples that typically come in the portable executable (PE) format (.EXE files) are constructed by a sequence of bits. Each malware binary consists of a string of zeros and ones. The zeros and ones can be converted or represented as an 8-bit vectors. The 8-bit vectors are then organized into two-dimensional matrices, which form greyscale images, a 2D matrix (corresponding to the height and the width of an image). Therefore, each grayscale pixel is represented by a value ranging from 0 to 255 [8].

Malware classification and detection by using image processing methodology was first proposed by Nataraj et al, who first converted malware binaries to grayscale images [24]. The premise behind converting malware into a greyscale image is to view malware from an image processing perspective. Image-based malware classification and detection aims to detect and classify the existence of a malware binary by studying the texture of the malware image, which is easily converted from the collected binary malware. This malware detection method easily lends itself to deep learning models, such as convolutional neural networks (CNN) that are widely used for image recognition. The beauty of this approach is that, unlike other malware detection approaches, it can even detect any small changes to malware code and, better yet, it can even detect packed as well as obfuscated malware. This is possible because when malware authors make changes to their code or pack their binaries or even obfuscate it, the texture will occur at a different position in the image representing malware [9]. When we convert malware samples into images, we can then apply deep learning algorithms to find visual patterns or similarities among malware families because malware authors often re-use code to create new variants. It's a known fact that malware is no longer been written but rather assembled [7].

A. NEED FOR THE STUDY

As mentioned earlier, machine learning (ML), particularly deep learning models, have been extensively applied to address many of the cyber security challenges, including malware detection. Many deep learning algorithms have been proposed over the past few years to address malware classification and detection. Such deep learning models rely on extracting important features in a process called "feature engineering". In essence, feature engineering allows researchers to select various features from both static as well as dynamic analysis of malware. Features corresponding to a particular class of malware are used to train a deep learning model in order create a separating plane between malware and clean ware [21]. Although previous research works [1, 6, 7, 21, 18,9] have made significant advances towards more efficient malware detection techniques using a variety of deep learning algorithms, the main issues is that most of such works develop an algorithm, apply it to a few datasets, achieve an acceptable level of accuracy, but do not strive to optimize their deep learning models. This is primarily because most cybersecurity researchers are not data scientists by trade and therefore, they lack the expertise to apply best practices to make their learning models optimized. We strongly believe that this creates a gap in the literature where there is lack of deep learning models with optimized characteristics. We believe that even if we develop a new learning algorithm, we should take the model to a new level and optimize it to make it more efficient and achieve higher accuracy-level results. The purpose of this research study is to bridge this gap in the literature by proposing a new convolutional deep learning neural network to detect malware accurately and effectively with high precision. This paper is different than most other papers in the literature in that it uses an expert data science approach by developing a convolutional neural network from scratch to establish a baseline of the performance model first, explores and implements an improvement model from the baseline model, and finally it evaluates the performance of the final model.

## B. MAJOR CONTRIBUTIONS OF THE STUDY

To fill the gap in literature, in this paper, a new convolutional deep learning network architecture for malware detection is proposed. The new CNN is based on an expert data science approach that develops a CNN from scratch and then uses data science best practices and approaches to optimize the model and ultimately achieve superior detection accuracy. Overall, the major contributions of our research work include the following:

1. We propose a new convolutional deep learning neural network to accurately and effectively detect malware with high precision.

2. We use an expert data science approach by developing a CNN from scratch to establish a baseline of the performance model first, explores and implements an improvement model from the baseline model,

3. We evaluate the performance of the final model. The baseline model initially achieves 98% accuracy rate but after increasing the depth of the CNN model, its accuracy reaches 99.183%.

4. Our novel deep neural network (DNN) model outperforms most of the CNN models in the literature.

5. Finally, to further solidify the effectiveness and accuracy of this CNN model, we use the improved model to make predictions on new malware samples within our dataset.

The rest of the paper is organized as follows. Section II reviews the related work in malware classification models considering various approaches traditionally employed for malware detection. In Section III, deep learning architecture is introduced to get insights into this research background. Section V presents the implementation architecture for deep learning in this research study and the statistical measures used to evaluate the performance of the classifier. Section V describes the methodology and dataset. Section VI discusses the experimental study and the results obtained for malware classification using our proposed CNN model. Section VII presents the comparison of our results with previous work.

Finally, Section IX provides the conclusion of this study and future work.

II.   RELATED WORK

To highlight the importance of our work, we investigated the application of machine learning and deep learning techniques for the detection and classification of malware by other researchers. Nataraj et al [1] are considered the pioneers in using machine learning for malware detection. They used a machine learning technique, k-NN, for their image-based malware classification. They used GIST descriptor to extract features from the input images. They used the much popular, but relatively small Malimg dataset with 9339 malware binaries from 25 different malware families. In their proposed model, they achieved an impressive 98% accuracy rate. One drawback for their novel approach was that the GIST descriptor is seen as time-consuming and overly complex.

In another relevant study, S. Yajamanam et al [10]. Took a slightly different approach by using the feature-enginering technique as an important factor to influence the accuracy of their malware classifier. So, they picked 60 features out of the 320 features available from GIST for training. Unfortunately, their model achieved only 92% accuracy because reducing features could have had a negative impact on the effectiveness and accuracy of their technique.

The authors in [11] used a different descriptor to extract features needed for pre-training their deep learning model of image recognition. They used ImageNet as their dataset which contained a whopping 1.2million images spanning 1,000 classes of malware binaries. Their thought was that the bigger the dataset is, the more effective and accurate the model would be. This was based on the premise that any given deep learning model is only as good as the training data. Much to their surprise, the proposed deep learning model achieved only 92% accuracy rate, which is not too impressive given the gigantic dataset they used and compared to other relevant models in the literature.

A lightweight approach was undertaken by Jiawei et al [25] for the detection of distributed denial of service malware. The researchers converted malware binaries into grayscale images and then fed them into fine-tuned CNNs. The machine learning classifier was used in local devices for further classification of the malware binaries. However, since the signature matching system contains details of each malware sample, its database is huge and therefore, is not efficient for IoT devices with limited resources. Once trained, only a small set of training data was required for the malware classification through machine learning, and therefore, a small two layer shallow CNN was used by the researchers for the malware detection. The researchers achieved an average accuracy rate of 94% on the classification of benign and malicious malware.

Unlike other previous studies from the literature, where image descriptors were used, Quan Le et al. [12] did not use descriptors, instead, they trained their deep learning model using input images. In their study, they converted raw input images into one-dimensional fixed size vectors before feeding into their CNN model. Although they achieved an impressive 98% accuracy rate with their proposed malware classifier, it must be noted that converting images to one-dimensional, fixed size vectors can negatively impact the quality of images, possibly leading to information loss.

In a similar vein, author of the paper [13] used raw malware images to train a CNN for malware classification. However, their model was flawed because they manipulated the data and balanced it in a way that provides higher accuracy rate.

Bensaud et al [24] used six deep learning models for the detection and classification of malware images and then compared the performance of these models with one another. The six models that the researchers ran on the Malimg dataset included Inception V3, VGG16-Net, ResNet50, CNN-SVM, MLP-SVM, and GRU-SVM. The researchers were unable to use grayscale images with the two models VGG16-Net and ResNet50, because the input layers of these images require the shape of 3, 224, 224, representing red, green, and blue (RGB) channels of the image, whereas the input layers in the grayscale images require 1, 224, 224 shapes. The figure below demonstrates the prediction accuracy of the six models:

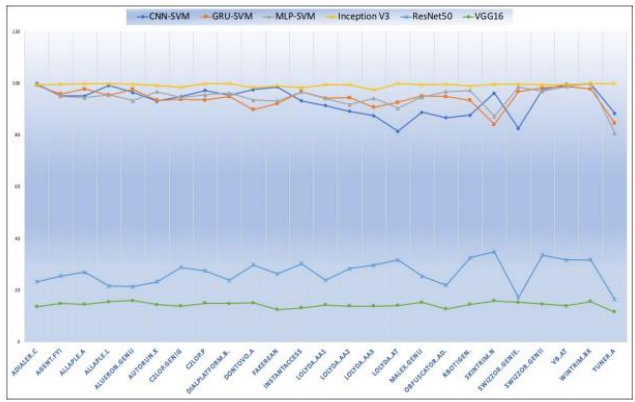

As can be seen, VGG16-Net and ResNet50 exhibited low performance in comparison with other models as these two models involved architectures that were designed to recognize colored images in RGB format. Therefore, these two models performed with the lowest accuracies when they were tested on the grayscale images. The highest accuracy of 99.24% was achieved by the Inception V3 model. CNN-SVM, GRU-SVM and MLP-SVM also performed well at 93.22%, 94.17%, and 94.55% accuracy rates respectively. But the accuracy rates of ResNet 50 and VGG16 were at 26.66% and 14.31% respectively.

Lad and Adamuthe (2020) proposed a combination of CNN and hybrid CNN + SVM model for malware image detection and classification. The researchers did not use Softmax as an activation function. They used SVM to perform the task of malware classification based on the features retrieved by the CNN model. The proposed model generated a vector of 256 neurons with the fully connected layer, which is input to SVM. The proposed model achieved 98.03% accuracy, which is better than the accuracy rates of other CNN models, such as Xception (97.56%), InceptionV3 (97.22%), and VGG16 (96.96%) [30].

It is becoming clear from the above-stated studies that it is not necessary to use image descriptors (e.g., GIST or SIFT [14]) to extract features for image processing, rather, we can simply utilize grayscale pixel values from malware images and use those values as features for training a machine learning model. The idea here is to reduce unnecessary complexity as well as time required to build an efficient deep learning malware classification model. It must be noted, though, that for this idea to work best, we need to ensure that the dimensions of a malware input-image should be around 32*32 so that we do not tax the malware classifier.

Inspired by aforementioned challenge, this paper investigates the impact of the dimension of the input malware images and the impact of the learning technique, i.e., CNN, applied on the performance of image-based malware classification. The experiment is conducted on Malimg dataset. We believe that even if we develop a new learning algorithm, we should take the model to a new level and optimize it to make it more efficient and achieve higher accuracy-level results. To this end, we aim to bridge this gap in the literature by proposing a new convolutional deep learning neural network to detect malware accurately and effectively with high precision.

## III.   SYSTEM ARCHITECTURE

This section introduces the CNN as a deep learning model to classify and detect image-based malware. As shown in Fig. 1, we convert malware executables into grayscale images and then group them into malware families such as Botnet, Banking Trojans, Backdoor, Worm and so on. Once this is done, we then proeed into feeding the image-based malware classifier, our CNN model.

The PE formats of the malware binaries are programs with name extensions, including .bin, .exe, and .dll [24]. PE files are identified based on their components, and these components are called .tex, .rdata, .data, and .rsrc. .tex is the first component that represents the code section and contains the instructions of the program. .rdata is the section that contains read only data, whereas .data contains modifiable data. .rsrc is the final component of PE file standing for resources used by the malware [24]. One of the simple methods for converting a binary file into a grayscale image is to treat the sequence bytes (8-bit group) of the malware binary as the pixel values of a gray- scale image (encoded by 8 bits) [main paper reference]. The figure below demonstrates the sections of a malware binary in a grayscale image composed of textural patterns [24]. Based on the textual patterns, malware can be classified.

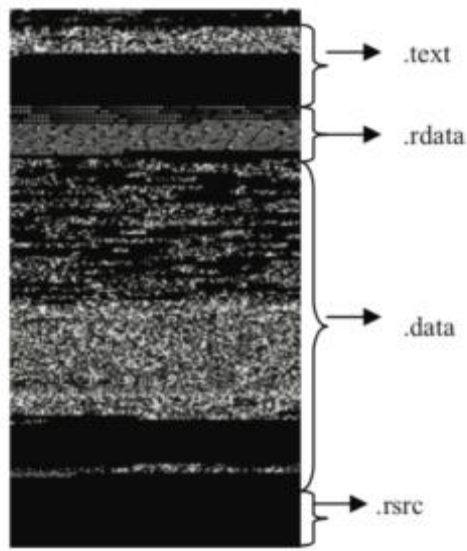

Visualization of malware binaries becomes better with images because the patterns hidden within such images become more pronounced and visible. As mentioned earlier, each grayscale pixel is represented by an 8 bit vector, the value of which can range from 00000000 (0) to 11111111 (255). Each 8-bit vector is represented by a number and can be converted into pixels in a malware image as shown below:

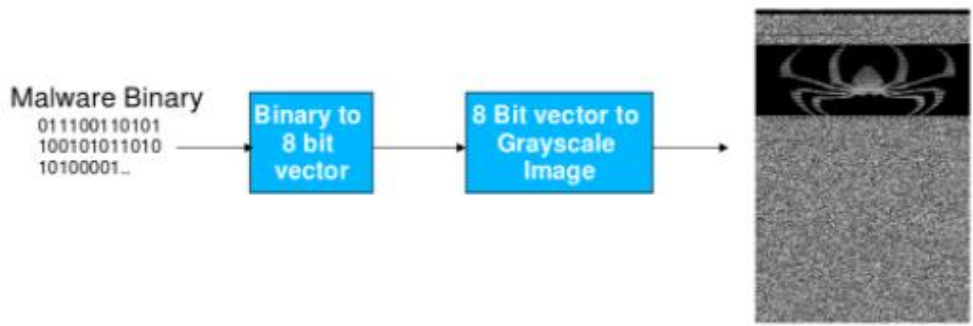

Fig. 1. Shows how to convert malware into an image.

Although the height of a malware-based image can vary depending on the size of the malware executable, its width is typically fixed size of 32,64, 128 pixesl. As the width of an image is usually fixed at 32, 64, 128 pixels [5] [13] [1

As a result, different malware binaries generate different malware images which have different shapes as illustrated in Fig. 2 (a), (b), and (c) for three malware families of Alueron.gen!J, Dialplatform.B, and Swizzor.gen!E respectively.

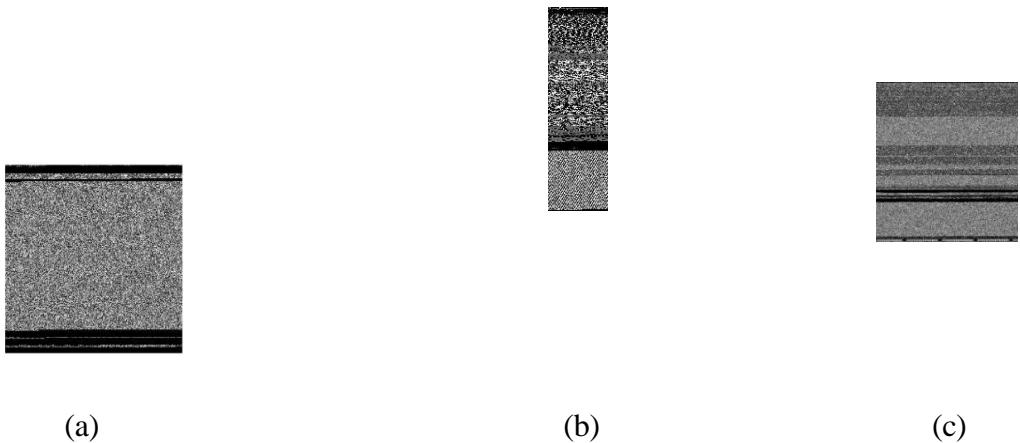

(a)                                    (b)                                    (c)

Fig. 2. Malware image of (a) Alueron.gen!J ; (b) Dialplatform.B; (c) Swizzor.gen!E

The advantage of using image-based malware classification models is that vairiants of the same malware family share very similar texture as illustrated in Fig. 3. Three variants of the Dontovo A family look very similar to the orgiginal malear. Those three variants were randomly selected among 431 variants stored in the Malimg dataset [5]. This similarity in texture will enable the CNN model to efficiently classify malware to their respective malware family based on the similarities in image texture.

Dontovo.A 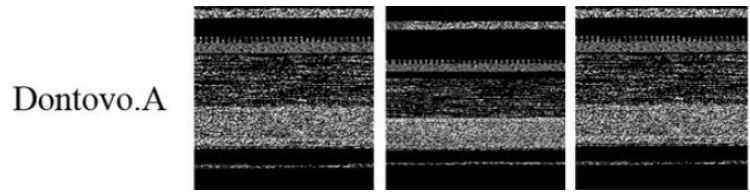

Fig. 3. Three variants of the Dontovo family

As previously highlighted, the machine learning models can be fed with grayscale malware images for training. To be precise, grayscale pixel values can be used as the features of the input images instead of extracted features from the use of image descriptors.

Original Dialplatform

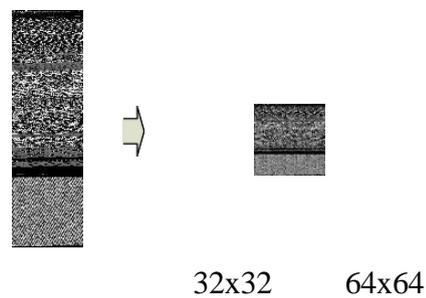

32x32        64x64

Fig. 4. The images of Dialplatform malware after normalization

CNNs refer to a subcategory of neural networks that can effectively classify and recognize specific features from images and therefore, CNNs are used extensively for visual image analysis. The application of CNN may range from recognition of images or videos, image classification, natural language processing [18], medical image analysis, and computer vision [6]. CNN has two primary functions, including feature extraction from images and classification of images [26]. The figure below illustrates the two functions of the CNN architecture:

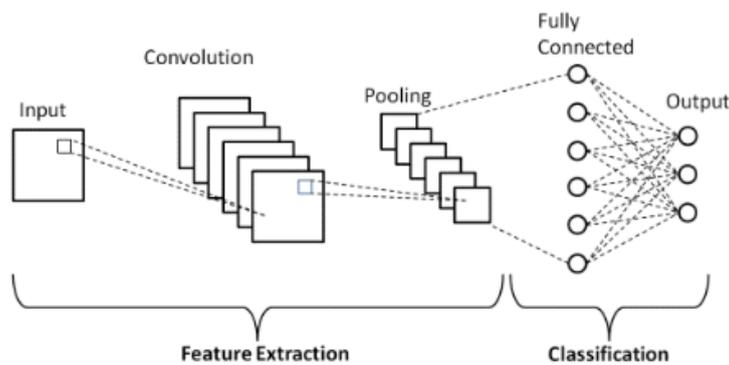

Two Functions of CNN [26]

The architecture of CNN primarily consists of two blocks [27]. The first block functions as a feature extractor by matching templates with the help of convolution filtering operations. The first layer is responsible for filtering images with many convolution kernels and produces feature maps, which are then resized or normalized. This process of filtering images and producing feature maps that are then normalized and resized is repeated several times [29]. The values derived from the last feature maps are finally used for concatenation into a vector. The output of the first block and the input of the second block are defined by this vector only. The figure below [29] demonstrates the first block, which is encircled in black:

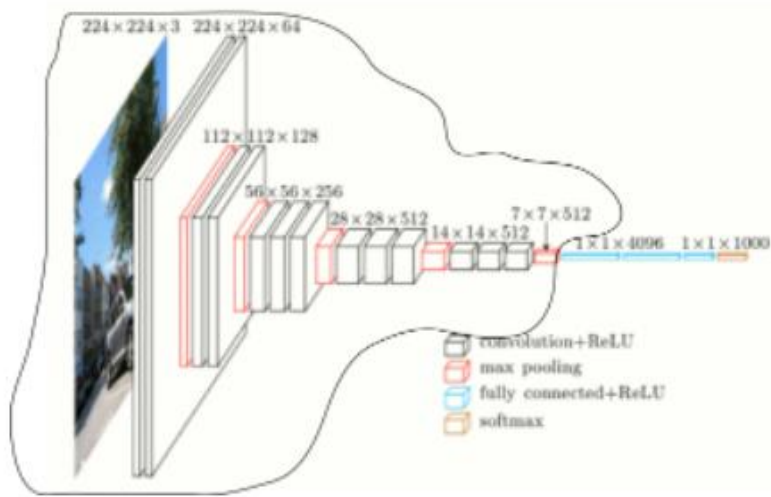

The function of the second block comes into the picture after all the neural networks are used for classification. The input factor values of the second block are transformed via several activation functions and linear combinations to generate a new vector to the output. The last vector has the same number of elements as classes. For example, element i represents the probability of the image belonging to class i. Each element has a value ranging between zero and one and the sum of all elements amounts to 1 [29]. The calculation of these probabilities is done by the last layer of the second block, which uses binary classification through a logistic function and multiclass classification through a softmax function as an activation function. As is the case with ordinary neural networks, gradient backpropagation determines the parameters of the layers. For example, during the training phase, the cross-entropy is minimized, but these parameters, in the case of CNN, refer specifically to the image features [29]. The figure below demonstrates the second block encircled in black:

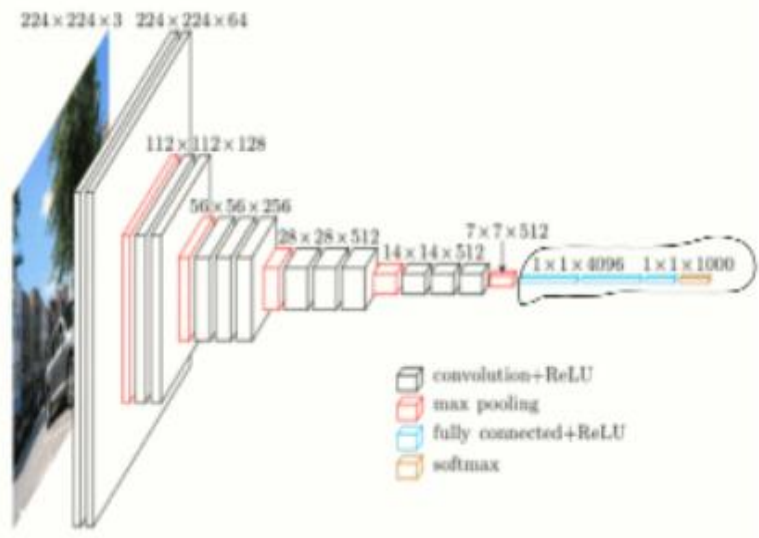

A CNN has four types of layers, including the convolutional layer, the pooling layer, the fully connected layer [19], and the ReLU correction layer [29].

The convolution layer is the first layer of CNN. The purpose of this layer is to detect the presence of a number of features in the images that are received as input [28]. The layer executes this purpose by convolution filtering. The principle behind the convolutional layer is to drag a window that represents the feature on an image and then estimate the convolution product laying between each segment of the scanned image and the feature itself. A feature is then viewed as a filter. Several images are received as input by the convolutional layer, and the convolution of each of these images is then calculated against each filter. The filters exactly represent the features we wish to see in the images. A feature map is obtained for each pair of image and filter, indicating the position of the features within the images [29]. The figure below demonstrates the convolution in which the central element of the kernel is positioned over the source pixel, which is then replaced with a weighted sum of itself and nearby pixels:

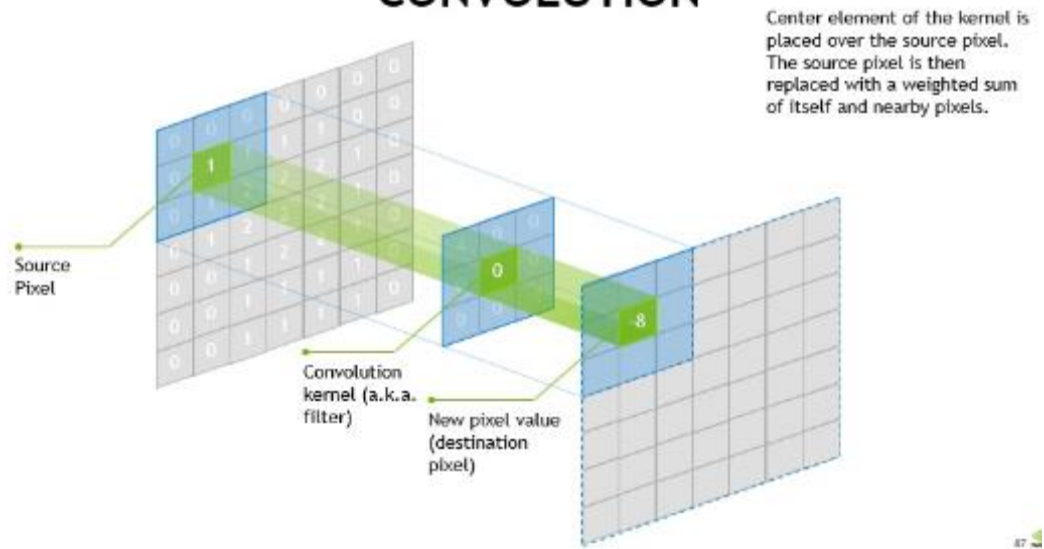

The positioning of the pooling layer is between two convolutional layers. The pooling layer receives many feature maps and then executes the pooling operations for each of them. The pooling operation involves minimizing the image sizes without affecting their important characteristics. The pooling layer achieves this by cutting the image into cells of regular sizes and maintaining the maximum value within each cell. The most common sizes of cells are 2 x 2 or 3 x 3 cells [29]. These cells remain separated from one another by a step of 2 pixels. By reducing the number of parameters and calculations done within the network, the pooling layer enhances the network efficiency by avoiding overlearning.

The ReLU correction layer stands for rectified linear units that refer to the real nonlinear function achieved by the application of the formula of ReLU(x) = max (0, x) [29]. The visual representation of the same is as below:

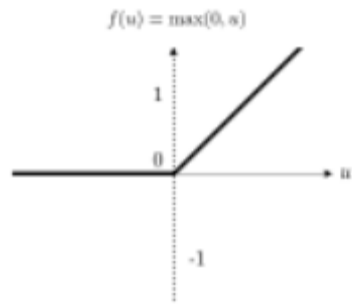

All negative values that are received as inputs by zeros are replaced by the ReLU correction layer. This layer performs the activation function.

The last layer of a CNN is the fully connected layer, which receives an input vector and generates a new output vector by applying a linear combination and an activation function for the input values received. This layer does the image classification as an input to the network and then generates a vector of size N, in which N represents the number of image classes. Each element of the vector represents the probability for the input image belonging to a class. The calculation of the probabilities is achieved by this layer by multiplying each input element by weight, then making the sum and applying an activation function. The relationship between the placement of features in an email and its class is determined by the fully connected layer. The input table of this layer is the output of the previous layer. Therefore, the input table corresponds to a feature map for a specific feature. The high values of an input table suggest the location of the feature within the image.

Table 1. Malimg dataset families [9]

| | Family Name | Variants |
|---|---|---|
| 1 | Allaple.L | 1591 |
| 2 | Allaple.A | 2949 |
| 3 | Yuner.A | 800 |
| 4 | Lolyda.AA 1 | 213 |
| 5 | Lolyda.AA 2 | 184 |
| 6 | Lolyda.AA 3 | 123 |
| 7 | C2Lop.P | 146 |
| 8 | C2Lop.gen!G | 200 |
| 9 | Instantaccess | 431 |
| 10 | Swizzor.gen!I | 132 |
| 11 | Swizzor.gen!E | 128 |
| 12 | VB.AT | 408 |
| 13 | Fakerean | 381 |
| 14 | Alueron.gen!J | 198 |
| 15 | Malex.gen!J | 136 |
| 16 | Lolyda.AT | 159 |
| 17 | Adialer.C | 125 |
| 18 | Wintrim.BX | 97 |
| 19 | Dialplatform.B | 177 |
| 20 | Dontovo.A | 162 |
| 21 | Obfuscator.AD | 142 |
| 22 | Agent.FYI | 116 |
| 23 | Autorun.K | 106 |
| 24 | Rbot!gen | 158 |
| 25 | Skintrim.N | 80 |
| | **Total** | **9,339** |

So, the overall CNN architecture will be as in the following diagram:

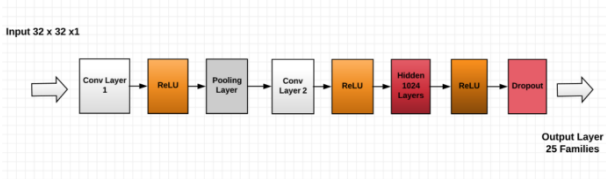

Fig. 5:  Overall CNN model architecture

IV.    METHODOLOGY AND DATASET

The Malimg dataset has been widely used in many research projects and experiments over the past few years as it certainly lends itself well to a good deep learning convolutional neural network. Something unique about this study is that the researchers decided to develop a new CNN model from scratch instead of reviewing the literature on well-performing models. The other unique aspect here is that the researchers, unlike most studies in the literature, did not just develop a model and presented the result, instead, they go above and beyond the normal expectations by developing a CNN model from scratch, develop a robust performance evaluation of the baseline model, explore the extensions to the baseline model in-order to improve the learning capacity of the model, and finally, the researchers will develop a finalized CNN model, evaluate the finalized model, and use it to make predictions on new malware.

The Malimg dataset perfectly lends itself to our CNN model as it has a good number of train and test dataset that we can use. To gain insight into the performance of our CNN model, and the learning curve, for a given training run, we can further break up the training dataset into a train and validation dataset. We use the traditional concept of 70% of the dataset for training and the remaining 30% for testing [22].

A.  DEVELOPING A BASELINE MODEL

In order to develop a baseline CNN model for our malware classification and detection task, we will begin by developing the infrastructure for the test harness which will enable us to evaluate our model on the Malimg dataset. This step is also important as it will establish the model baseline, evaluate it, and ultimately improve the model. Developing the infrastructure for our test hardness involves the following steps: load the Malimg dataset, prepare the dataset, define our CNN model, evaluate the model, and finally, make use the model to make predictions using new malware samples (the hold-out test) of out Malimg dataset.

## B. LOADING THE DATASET

We know some things about the dataset. The images all have the same square size of 32×32 pixels, and that the images are grayscale. Therefore, we can load the images and reshape the data arrays to have a single-color channel. The following code snippet illustrates how to achieve the above task:

```
# load dataset
(trainX, trainY), (testX, testY) = malmig.load_data()
# reshape dataset to have a single channel
trainX = trainX.reshape((trainX.shape[0], 32, 32, 1))
testX = testX.reshape((testX.shape[0], 32, 32 1))
```

## C. DEFINING OUR PROPOSED CNN MODEL

Now, we need to define a baseline convolutional neural network model for malware detection and classification. The model has two main aspects: the feature extraction front end comprised of convolutional and pooling layers, and the classifier backend that will make a prediction to determine if a binary sample is malware or not.

We will deploy a conservative setup for the stochastic gradient descent optimizer with a learning rate of 0.01 and a momentum of 0.9. The categorical cross-entropy loss function will be optimized, suitable for multi-class classification, and we will monitor the classification accuracy metric (machinelearningmastery, 2021).

Our model can be defined using the following code:

```
# define cnn model
def define_model():
        model = Sequential()
        model.add(Conv2D(32, (3, 3), activation='relu', kernel_initializer='he_uniform', input_shape=(28, 28, 1)))
        model.add(MaxPooling2D((2, 2)))
        model.add(Flatten())
        model.add(Dense(100, activation='relu', kernel_initializer='he_uniform'))
        model.add(Dense(10, activation='softmax'))
        # compile model
        opt = SGD(lr=0.01, momentum=0.9)
        model.compile(optimizer=opt, loss='categorical_crossentropy', metrics=['accuracy'])
        return model
```

## C. EVALUATING OUR CNN MODEL

Once we have completed the task of defining our CNN model for malware detection, we will now need to evaluate it. We will evaluate the model using five-fold cross-validation. The value of k=5 will ensure that we obtain a baseline for both repeated evaluation and would also ensure a relatively short running time. So, the k=5 means that our training dataset will be split into 5 test sets (machinelearningmastery, 2021).

Additionally, to ensure that our model will contain the same train and test datasets in each of the five folds mentioned above, we will be shuffling the training dataset before starting the split process (machinelearningmastery, 2021). This process will ensure that we are comparing "apples-to-apples" (for fair model comparison) so to speak. The batch size for training our CNN baseline model will be 32 malware samples with 10 training epochs. This setup will enable us to estimate the performance of our CNN baseline model, track the result history of each run, and malware classification accuracy of each of the 5 folds.

The following code illustrates how to achieve the above task:

```
# evaluate a model using k-fold cross-validation
def evaluate_model(dataX, dataY, n_folds=5):
        scores, histories = list(), list()
        # prepare cross validation
        kfold = KFold(n_folds, shuffle=True, random_state=1)
        # enumerate splits
        for train_ix, test_ix in kfold.split(dataX):
                # define model
                model = define_model()
                # select rows for train and test
    trainX, trainY, testX, testY = dataX[train_ix], dataY[train_ix], dataX[test_ix],          dataY[test_ix]
                # fit model
```

```
history = model.fit(trainX, trainY, epochs=10, batch_size=32, validation_data=(testX, testY),
                    verbose=0)

    # evaluate model

    _, acc = model.evaluate(testX, testY, verbose=0)

    print('> %.3f' % (acc * 100.0))

    # stores scores

    scores.append(acc)

    histories.append(history)

return scores, histories
```

## V.   EMPIRICAL RESULTS

The next logical step after evaluating our model is to display and present the results. We will focus on the learning behavior of our CNN model and then estimate its performance. As its commonly known in the deep learning community, over-fitting and under-fitting are two major issues associated with virtually any deep learning model. To this end, and to ensure that our CNN model is not flawed with neither over-fitting nor under-fitting, we have created a line-plot to display and report the performance of our model on the test as well as train dataset during each of the five-fold cross-validation. The line-plots will provide insight into whether or not our model is under-fitted or over-fitted for the malware dataset [13].

Only one figure with two subplots is needed to gain the insight into the over-fitting and under-fitting issue, one subplot for loss and one for accuracy of our CNN model (machinelearningmastery, 2021). The performance of our model on the training dataset is shown using the blue lines while the orange lines will display our CNN model performance on the hold out test dataset. The code-snippet below will achieve this task:

# summarize model performance

```
def summarize_performance(scores):

        # print summary

        print('Accuracy: mean=%.3f std=%.3f, n=%d' % (mean(scores)*100, std(scores)*100,

len(scores)))

        # box and whisker plots of results

        pyplot.boxplot(scores)

        pyplot.show()
```

We can see these cases where the model achieves perfect skill. These are good results.

> 98.467

> 98.683

> 98.642

> 98.850

> 98.583

The numbers above are impressive and they clearly show that our model evaluation is progressing.

Next, a diagnostic plot is shown, giving insight into the learning behavior of the model across each fold.

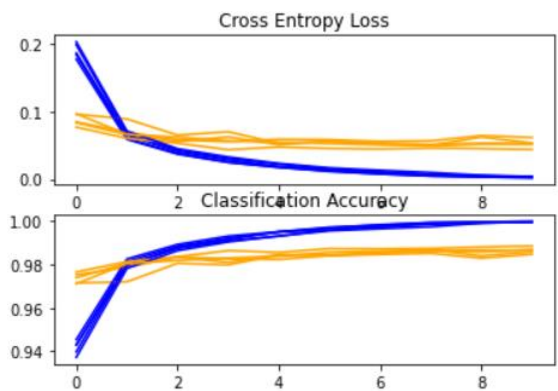

Fig 6: - Loss and Accuracy Learning Curves for the Baseline Model During k-Fold Cross-Validation

In this case, we can see that the model generally achieves a good fit, with train and test learning curves converging. There is no obvious sign of over- or underfitting.

Next, a summary of the model performance is calculated. We can see in this case, the model has an estimated skill of about 98.6%, which is reasonable.

Finally, a box and whisker plot are created to summarize the distribution of accuracy scores.

A. IMPROVING THE BASELINE MODEL

There are many ways that we might explore improvements to the baseline model. We can explore aspects of our CNN model configuration that are likely to result in an improvement, so-called low-hanging fruit. The first is a change to the learning algorithm, and the second is an increase in the depth of the model ((Machine learning mastery, 2021)).

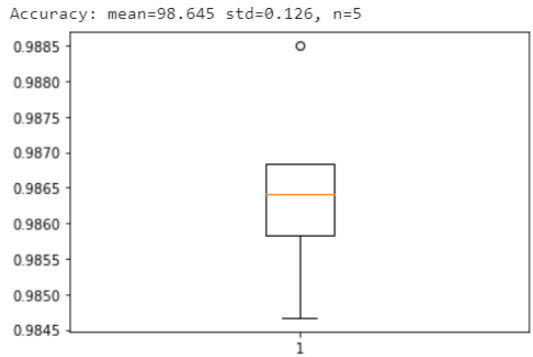

Figure 7. Box and Whisker Plot of Accuracy Scores for the Baseline Model Evaluated Using k-Fold Cross-Validation

Now we can explore a couple ways to make improvements to our CNN model. We can change the model configuration to explore improvements over the baseline model. Two common approaches involve changing the capacity of the feature extraction part of the model or changing the capacity or function of the classifier part of the model (Machine learning mastery, 2021).

We can increase the depth of the feature extractor part of the model, following a VGG-like pattern of adding more convolutional and pooling layers with the same sized filter, while increasing the number of filters. In this case, we will add a double convolutional layer with 64 filters each, followed by another max pooling layer ((Machine learning mastery, 2021)).

Running the example reports model performance for each fold of the cross-validation process as shown below:

> 98.775

> 98.683

> 98.967

> 99.183

> 99.008

The per-fold scores above may suggest some improvement over the baseline.

B. FINALIZING OUR MODEL AND MAKING PREDICTIONS

Although it appears intriguing to continue making improvements on our model, at this point we will choose the final configuration of our CNN model. The final version of our model will be the deeper model. We have finalized our model by fitting it on the entire training Malimg training dataset and then load the model and evaluate it. We will evaluate our model performance and accuracy on the hold out test Malimg dataset. This will give us insight into how practically accurate our model performs on real malware dataset.

The classification accuracy for the model on the test dataset is calculated and printed. In this case, we can see that the model achieved an accuracy of 99.180%, or just less than 1%, which is not bad at all with a standard deviation of about half a percent (e.g. 99% of scores).

> 99.180

| Year | Researchers | Methods | Technique | Accuracy (%) |
|------|-------------|---------|-----------|--------------|
| 2011 | Nataraj et al. | GIST | Machine Learning | 98 |
| 2017 | S. Yue | CNN | Deep Learning | 97.32 |
| 2017 | Makandar and Patrot | Gabor wavelet-kNN | Machine Learning | 89.11 |
| 2018 | Yajamanam et al. | GIST+kNN+SVM | Machine Learning | 97 |
| 2018 | Cui, Xue, et al. | GIST+SVM [15] | Deep Learning | 92.20 |
| 2018 | Cui, Xue, et al. | GIST+kNN | Deep Learning | 91.90 |
| 2018 | Cui, Xue, et al. | GLCM+SVM | Deep Learning | 93.20 |
| 2018 | Cui, Xue, et al. | GLCM+kNN | Deep Learning | 92.50 |
| 2018 | Cui, Xue, et al. | IDA+DRBA | Deep Learning | 94.50 |
| 2019 | Cui, Du, et al. | CNN, NSGA-II | Deep Learning | 97.6 |
| 2020 | Mallet | CNN, Keras | Deep Learning | 95.15 |
| 2020 | Vasan et al. | IMCFN, Color images | Deep Learning | 98.82 |
| 2021 | Moussas& Andreatos | Image and file features, ANN | Two-level ANN | 99.13 |
| 2021 | Omar & Shiaeles | CNN, Keras | Deep Learning | 99.18 |

Table 3. Comparisons of the proposed DL-CNN with other existing deep learning approaches.

## VI. RESULTS COMPARISON WITH PREVIOUS WORK

To ensure fairness of comparison (comparing apples-to-apples) we provide a comparison of our deep learning CNN model's performance with other deep learning algorithms for malware classification and detection using the same dataset. The final version of our proposed convolutional deep learning model achieved an impressive accuracy score of 99.18 compared to Mallet (2020) which has 95.15. This clearly shows our model is outperforming other models in the literature. It's interesting to note that Mallet (2020) used almost an identical deep learning architecture that involved CNN and the Keras scientific environment. Also, our model outperforms the deep learning model proposed by Cui, Du, et al (2019) which achieved an accuracy score of 97.6. It is also interesting to note that our model even outperforms a recent work conducted by Moussas&Andreatos (2021) with an impressive accuracy rate of 99.13 on the exact same dataset.

Table 3. Comparisons of the proposed DL-CNN with other existing deep learning approaches.

## VII. CONCLUSION

In this paper, we proposed, developed, and presented a deep learning convolutional neural network modelfor malware detection. Our model deployes a unique approach to malware detection in that it developed a deep learning CNN from scratch, developed a bseline model, developed a robust performance evaluation of the baseline model, explored the extensions to the baseline model in-order to improve the learning capacity of the model, and finally, we developed a finalized CNN model, evaluated the finalized model, and used it to make predictions on new malware.

We evaluated our model using the popular Malimg dataset of 9339 samples belonging to twnety five malware families. Comparing our model's performance to existing DL-based frameworks, our results clearly illistrate that our deep learning CNN outpeforms those works presented in existing deep learning-based malware detection models.

As future work, we believe that the perrformance of our model can be further tested and evluated using biger datasets such as the BIG 2015 dataset and newly made available datasets from repositoies such as Kaggle. Another future direction would be to compare our model's accuracy score to the accuracy score of other similar deep learning models such as the k-NN model and others.

Sec. and Privacy, 719-726 (2015).

43, (2020). DOI: 10.5815/ijcnis.2020.06.03

31. Lee, C. et al. An Evaluation of Image-Based Malware Classification Using Machine Learning. In

   book: Advances in Computational Collective Intelligence, 12th International Conference, ICCCI

   2020, Da Nang, Vietnam, November 30 – December 3, 2020, Proceedings. (2020). DOI:

   10.1007/978-3-030-63119-2_11

Omar et al. (2023) "**Text-Defend: Detecting Adversarial Examples using
   Local Outlier Factor" !7th IEEE International conference of computing semantics**
Cybercrime and the Nature of Insider Threat Complexities in Healthcare and Biotechnology Engineering
   Organizations
DN Burrell, C Nobles, A Cusak, M Omar, L Gillesania - Journal of Crime and Criminal Behavior, 2022

Omar, M. et al. (2022). "Robust Natural Language Processing: Recent Advances, Challenges, and
   Future Directions". Under review with *Proceedings of the IEEE Journal*.
Omar, M., Jamal Al Karaki., et al. (2022). "Reverse-Engineering Malware".  Book chapter to appear in
"Cybersecurity Capabilities in Developing Nations and Its Impact on Cybersecurity". IGI Global.
DOI: 10.4018/978-1-7998-8693-8

Alkinoon, Omar, et al.  (2021). "Security Breaches in the Healthcare Domain". CoSNet 2021 Conference.
Omar (2021). "Developing Cybersecurity Education Capabilities at Iraqi Universities" Proceedings of
   the Sixteenth Midwest Association for Information Systems Conference, Peoria, Illinois, May
   20-21, 2021

Dawson, M., & **Omar, M**. (2015) "*Handbook of Research on New Threats and Countermeasures in
Digital Crime and Cyber Terrorism".* (Advances in Information Security, Privacy and Ethics Book
Series) (pp.1-305) Hershy, PA: IGI global  Publishing.
Dawson, M. Eltayeb. M. & **Omar, M.** (2016) "*Security Solutions for Hyper connectivity and the
Internet of Things".*   (Advances in Information Security, Privacy and Ethics Book Series) (pp.1-305)
Hershy, PA: IGI global  Publishing.

Dawson, **M., Omar**, M., Abramson, J., Leonard, B., & Bessette, D. (2017). Battlefield Cyberspace:
Exploitation of Hyperconnectivity and Internet of Things. In M. Dawson, D. Kisku, P. Gupta, J. Sing, &
W. Li (Eds.) Developing Next-Generation Countermeasures for Homeland Security Threat
Prevention (pp. 204-235). Hershey, PA: Information Science Reference. doi:10.4018/978-1-5225-
0703-1.ch010
Dawson, M., Wright, J., **Omar, M**., & Leonard, B. (2015) Mobile Devices: The Case for Security
Hardened Systems. *Handbook of Research on New Threats and Countermeasures in Digital Crime
and Cyber Terrorism.* (Advances in Information Security, Privacy and Ethics Book Series) (pp.1-
305) Hershy, PA: IGI global  Publishing
.**Omar, M**., (2015)  "Insider Threats". *Handbook of Research on New Threats and Countermeasures
in Digital Crime and Cyber Terrorism.*  (Advances in Information Security, Privacy and Ethics Book
Series) (pp.1-305) Hershy, PA: IGI global  Publishing.

.